\documentclass[5p]{elsarticle}

\usepackage{amssymb}
\usepackage{amsmath,amssymb,amsfonts}
\usepackage{algorithmic}
\usepackage{algorithm}
\usepackage{algorithmic}
\usepackage{graphicx}
\usepackage[justification=raggedright]{caption} 
\usepackage{multirow}

\journal{arXiv}

\begin{document}
\begin{sloppypar}
\begin{frontmatter}



\title{Enhancing Nucleus Segmentation with HARU-Net: A Hybrid Attention Based Residual U-Blocks Network}

\author[label1]{Junzhou Chen}
\affiliation[label1]{organization={Hohai University},
            city={Nanjing},
            postcode={211100},
            state={China},
            country={junzhouchen7@gamil.com}}

\author[label2]{Qian Huang}
\affiliation[label2]{organization={Hohai University},
            city={Nanjing},
            postcode={211100},
            state={China},
            country={huangqian@hhu.edu.cn}}


\author{Yulin Chen, Linyi Qian, Chengyuan Yu}




\begin{abstract}
Nucleus image segmentation is a crucial step in the analysis, pathological diagnosis,  and classification, which heavily  relies on the quality of nucleus segmentation. However, the complexity of issues such as variations in nucleus size, blurred nucleus contours, uneven staining, cell clustering, and overlapping cells poses significant challenges. Current methods for nucleus segmentation primarily rely on nuclear morphology or contour-based approaches. Nuclear morphology-based methods exhibit limited generalization ability and struggle to effectively predict irregular-shaped nuclei, while contour-based extraction methods face challenges in accurately segmenting overlapping nuclei. To address the aforementioned issues, we propose a dual-branch network using hybrid attention based residual U-blocks for nucleus instance segmentation. The network simultaneously predicts target information and target contours. Additionally, we introduce a post-processing method that combines the target information and target contours to distinguish overlapping nuclei and generate an instance segmentation image. Within the network, we propose a context fusion block (CF-block) that effectively extracts and merges contextual information from the network. Extensive quantitative evaluations are conducted to assess the performance of our method. Experimental results demonstrate the superior performance of the proposed method compared to state-of-the-art approaches on the BNS, MoNuSeg, CoNSeg, and CPM-17 datasets. 
\end{abstract}




\end{frontmatter}


\section{Sample Section Title}
\label{sec:sample1}

H\&E (Hematoxylin and Eosin) staining of tissue sections is a widely used histological staining technique that allows for the visualization of cellular tissue structures and morphological features such as nucleus morphology. These features play a crucial role in the pathological diagnosis and the development of treatment plans. Nucleus image segmentation is a key step in pathological diagnosis, aiming to label all pixels corresponding to individual nuclei in histopathological images \cite{b1} \cite{b2}. The analysis, diagnosis, classification, and grading of cancer heavily rely on the quality of nucleus segmentation \cite{b3}. However, traditional manual screening methods are time-consuming and labor-intensive, with the quality of results highly dependent on the personal experience and condition of the screeners, leading to potential human errors \cite{b4}. Consequently, in recent years, numerous studies have focused on utilizing machine-based screening approaches \cite{b5,b6} to assist experts in reducing errors and workload, accelerating the screening process, and lowering screening costs.

\begin{figure}[ht]

\centerline{\includegraphics[width=\columnwidth]{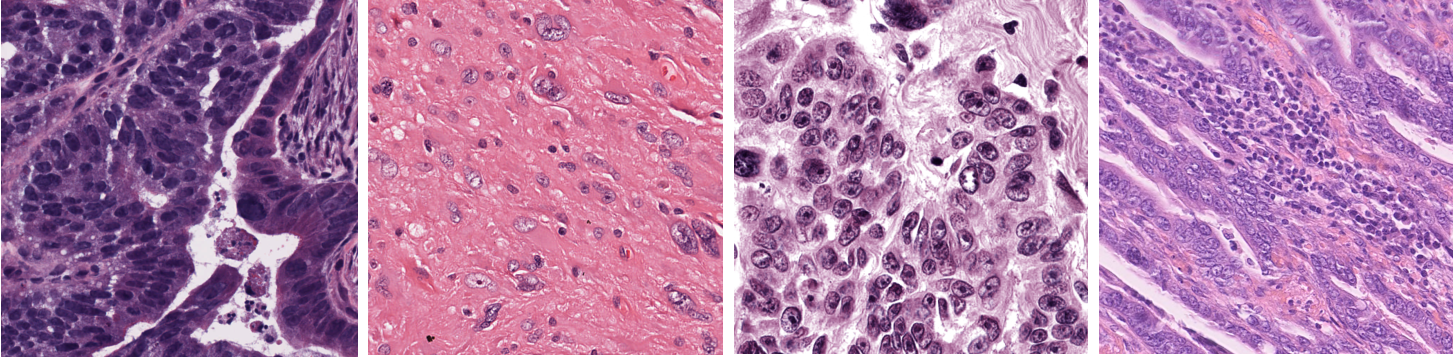}}
\caption{Nucleus images from different datasets.}
\label{fig1}
\end{figure}

Due to the imaging conditions and distribution characteristics of nuclei under the microscope, the segmentation of nuclei still presents many difficulties and challenges. In Fig. \ref{fig1}, the density distribution and adhesive edges of nuclei often result in over-segmentation or under-segmentation, thereby affecting the accuracy of the segmentation. Furthermore, complex factors such as varying sizes and shapes of nuclei, blurred nuclear contours, uneven staining, overlapping cell clusters, and numerous cells contribute to a high error rate in nucleus segmentation \cite{b7}. Regarding nuclear instance segmentation based on distance\cite{a1} \cite{a2}, such as StarDist\cite{stardist}, the centroid is often located far away from boundary pixels for large-sized nuclei, resulting in degraded distance prediction accuracy. Additionally, supervision is imposed on each respective distance value, and there is a lack of global constraint on the shape of each nucleus \cite{cppnet}. On the other hand, traditional contour-based methods for nucleus instance segmentation often yield poor segmentation results due to the fuzzy boundaries between nuclei \cite{a3} \cite{a4}. In-depth research and development of nucleus segmentation algorithms can significantly enhance the efficiency and accuracy of digital pathology\cite{b8}, providing a more scientific basis for the early diagnosis and treatment of diseases.

The existing methods for nucleus segmentation in H\&E-stained tissue slides have shown unsatisfactory results. The main reasons for this can be summarized as follows: (1) In complex cell images, nuclei are often mixed with other structures and noise, making it challenging to segment them accurately. This leads to increased difficulty in nucleus segmentation, consequently affecting the overall segmentation accuracy. (2) Insufficient attention has been given to the boundary regions of nuclei in existing research, which can result in errors when segmenting overlapping nuclei. (3) Existing instance segmentation methods often require accompanying object detection, such as using Mask-RCNN \cite{b9}, leading to overly complex models. (4) Nuclei exhibit diverse morphologies, and instance segmentation methods based solely on nuclear morphology often lack the required robustness.

The proposed research addresses the complex challenges in nucleus segmentation by introducing a novel neural network model called HARU-Net, which combines the U2Net architecture \cite{b10} with a hybrid attention mechanism \cite{b11}. Building upon U2Net, HARU-Net incorporates attention mechanisms into Residual U-blocks to handle these challenging images effectively. The HARU-Net model adds a context encoding layer to learn contextual features and incorporates hybrid attention learning modules in each layer to focus on relevant regions of the features. Additionally, HARU-Net predicts both foreground and contour images of nuclei and combines them to obtain instance-level nucleus segmentation results. Compared to traditional networks \cite{b14}\cite{b13}, HARU-Net better captures detailed information in the images, leading to improved segmentation accuracy and robustness. Experimental evaluations conducted on four publicly available datasets (BNS\cite{BNS}, MoNuSeg\cite{MoNuSeg}, CoNSep \cite{b12} and CPM-17\cite{cpm}) demonstrate that HARU-Net achieves state-of-the-art performance. Therefore, the HARU-Net network model can effectively be applied to nucleus segmentation tasks.

The main innovations and contributions of this article are as follows:
\begin{itemize}
\item We design a dual-task network for nucleus segmentation. By combining the hybrid attention mechanism with nested Residual U-blocks and utilizing an improved context for task prediction, we achieved significant performance improvements in the field of nucleus segmentation.

\item We propose a novel context fusion mechanism that significantly enhances the model's performance.

\item A simple and effective method is introduced to convert semantic segmentation information into instance-level information.

\item The HARU-Net are evaluated on different datasets, including TNBC, MoNuSeg, CoNSep, and CPM-17 datasets. In the experiments, HARU-Net outperformed the state-of-the-art models in terms of performance.
\end{itemize}

\begin{figure*}[!t]
\centerline{\includegraphics[width=\textwidth]{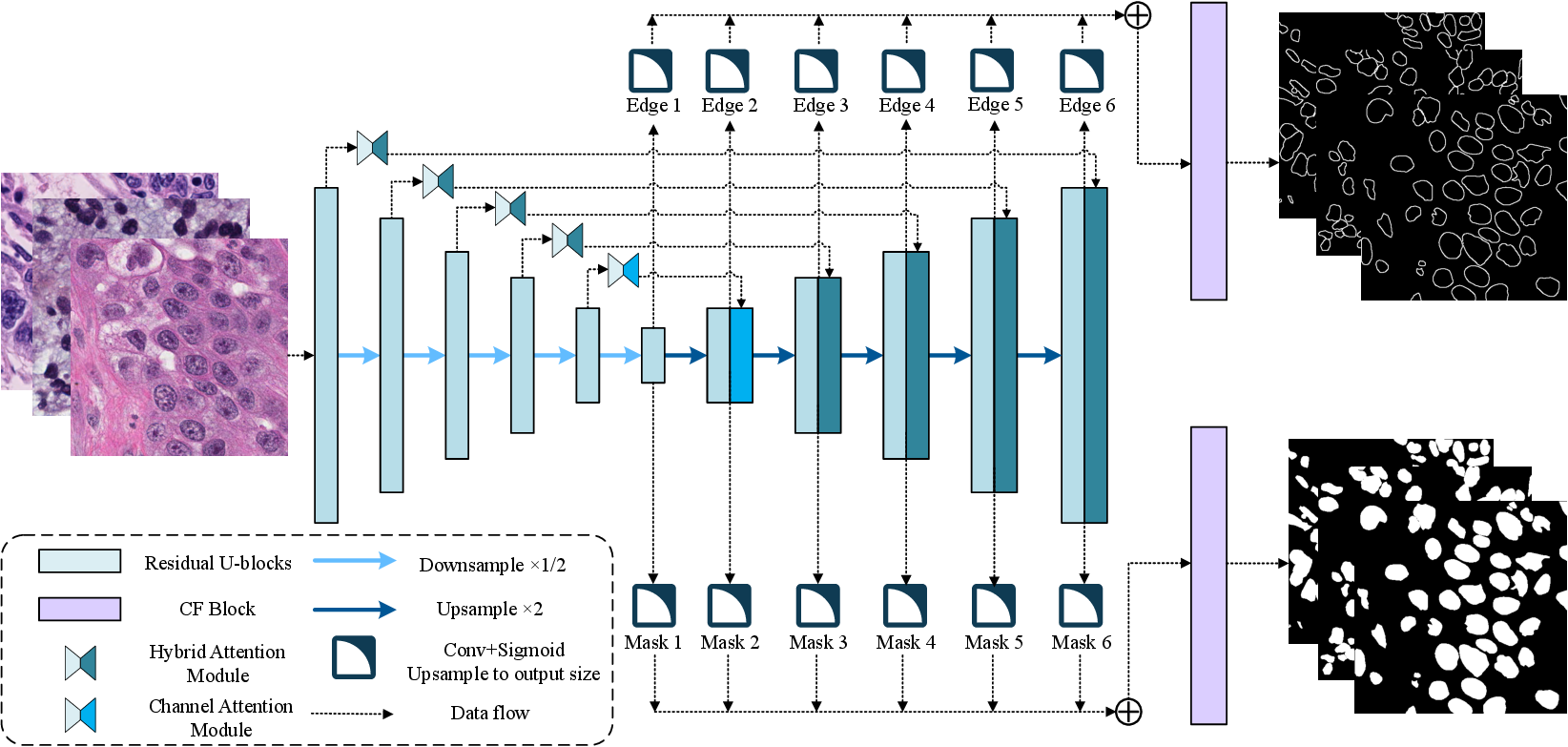}}
\caption{Overall architecture of the network demonstrates the precise placement of modules within the network. The channel attention mechanism is applied at the deepest layer, while the remaining layers utilize a hybrid attention mechanism.}
\label{fig2}
\end{figure*}

\section{Related work}

Currently, methods for nucleus image segmentation can be classified into traditional methods\cite{b18} and deep learning-based methods. Traditional segmentation methods include threshold-based segmentation \cite{b19}, region-based segmentation\cite{b20}, graph-based segmentation\cite{b21}, superpixel-based segmentation\cite{b22}, fuzzy clustering, and other methods\cite{b23}. These methods construct image segmentation models by manually selecting features, which results in good performance only on specific feature-rich datasets or samples. For example, C.H.Lin et al.\cite{b24}used a segmentation method based on a series of edge enhancement techniques, but it performed poorly on blurry nuclear contours. Yang Song et al.\cite{b25}employed a contrast-based adaptive version of the mean shift and SLIC algorithms, along with intensity-weighted adaptive thresholds, for segmenting nuclei in Papanicolaou (Pap) smear images. However, in many cases, the aforementioned traditional methods struggle to handle cervical cell images with irregular shapes and sizes. To address this issue, deep learning-based methods have gradually become mainstream. In recent years, research on image segmentation using convolutional neural networks (CNNs)\cite{b26}\cite{b27} has achieved remarkable results.

With the continuous improvement of computing performance, deep learning algorithms have demonstrated outstanding performance in image segmentation. In the field of nuclear image segmentation, the most commonly used network architectures include FCN\cite{b28}, Mask R-CNN\cite{b9}, U-Net\cite{b29}, and others. In particular, U-Net was initially applied to medical image segmentation tasks, utilizing skip connections to connect the intermediate downsample and upsample layers to extract contextual information. However, due to the direct fusion of low-level and high-level features through skip connections, there may be semantic gaps and difficulties in handling overlapping regions. To address this issue, subsequent researchers have proposed methods such as Attention U-Net\cite{b30}, U2Net\cite{b31}, CE-Net\cite{b31}, CIA-Net\cite{b32}, AL-Net\cite{b33}, and others. These methods have demonstrated good performance on nuclear segmentation datasets. Among them, CE-Net\cite{b31} extends the application of U-Net in medical image segmentation by incorporating enhanced network structures such as DAC and RMP blocks. To obtain more advanced information while preserving spatial details, the Context Encoding Network (CE-Net) replaces the encoding module of U-Net with a pre-trained model, resulting in improved performance on 2D medical image segmentation tasks. The Attention U-Net\cite{b30} model is a U-Net model that incorporates attention mechanisms. This model has the ability to automatically learn the shape and size of the targets, significantly improving the sensitivity and accuracy of the model with minimal additional computational cost. Similarly, Xuebin et al.\cite{b31} designed a two-level nested U-shaped structure called U2Net for salient object detection (SOD). U2Net captures more contextual information from different scales, leading to better performance in SOD tasks.

Attention mechanisms are also crucial for nuclear segmentation. The essence of attention mechanisms is to weigh different elements by multiplying each element in a matrix by its corresponding weight \cite{b34}. Attention mechanisms can extract nonlinear features and simplify computations. Hu et al. \cite{b35} introduced a compact module to leverage relationships between channels and establish a channel attention module. In spatial attention mechanisms, the weight for each pixel in the image is computed \cite{b36}. In the hybrid attention mechanism, Jing Zhao et al. \cite{b37} combined spatial attention and channel attention to design CBAM, which can learn features at different levels and weight them for fusion, achieving excellent results in the visual domain. In this paper, we improve upon the CBAM attention mechanism and combine it with a two-level nested U-shaped structure to develop a more practical and effective network model.

In the context of nuclear segmentation and the need, for instance, delineation, DCAN \cite{b38} proposed a novel Deep Contour-Aware Network that simultaneously segments nuclei and their boundaries. The multi-task segmentation framework in DCAN has been widely used in nuclear segmentation methods. To capture multi-scale spatial information, a Spatial Perception Network (SpaNet) \cite{b39} was proposed. SpaNet's multi-scale dense units are equipped with a feature aggregation property that allows positional information to flow throughout the network. In the aforementioned methods, there is no feature fusion between the multi-task branches. An information aggregation module is introduced to fuse feature maps from different branches. Building upon this, a Boundary-assisted Region Proposal Network (BRP-Net) \cite{b40} is proposed. HoVer-Net \cite{b12} is another method proposed for simultaneous nucleus segmentation and classification. In this network, distance information between nucleus pixels and their centroids is introduced in both vertical and horizontal directions. These distance features are utilized to help the network learn information about nucleus shape and structure, thereby improving segmentation accuracy.

\section{Method}
\label{sec:Method}

Our network model's overall architecture is shown in Fig. \ref{fig2}. The module details within the network are illustrated in Fig. \ref{fig3} to Fig. \ref{fig5}, and the instance segmentation process is depicted in Fig. \ref{fig6}. We first utilize a dual-branch network to detect nucleus pixels and contours. Finally, by combining the semantic information of the contours and nuclei, we perform instance segmentation.

\subsection{Residual U-blocks}

\begin{figure}[ht]
\centerline{\includegraphics[width=0.7\linewidth]{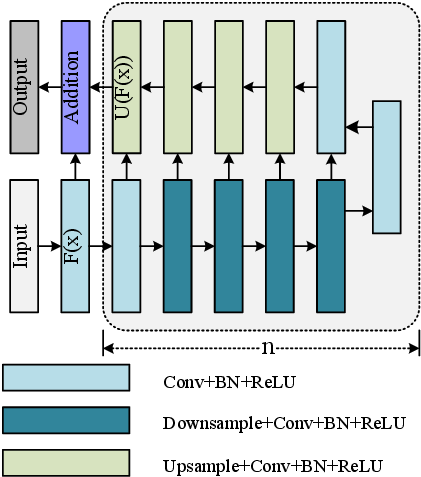}}
\caption{The Residual U-Block (RSU) structure is illustrated, where the value of 'n' decreases by 1 with the increasing depth of the network, starting from 7.}
\label{fig3}
\end{figure}

\begin{figure*}[t]
\centerline{\includegraphics[width=\textwidth]{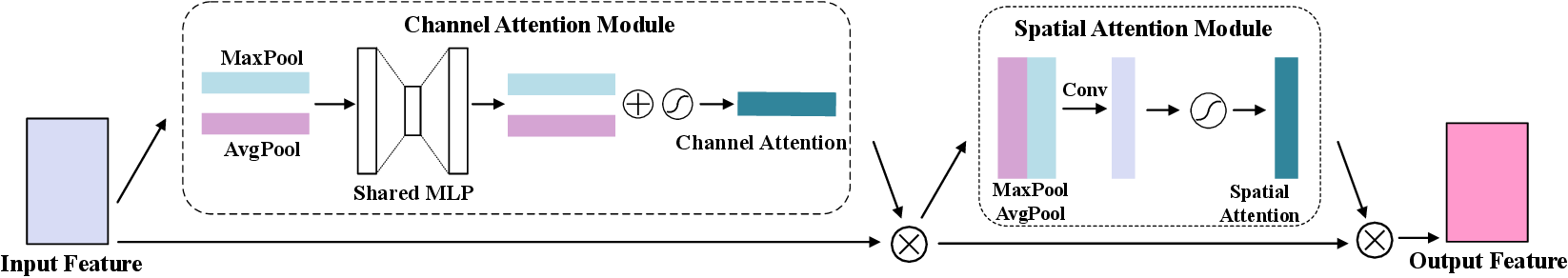}}
\caption{CBAM attention mechanism applies channel attention weighting and spatial attention weighting to the input sequentially.}
\label{fig4}
\end{figure*}
Fig. \ref{fig3} illustrates the Residual U-blocks, which are a novel network structure proposed in the U2Net\cite{b10} paper to capture multi-scale features within a single stage. It is an improvement based on the U-Net network structure and embeds the U-Net structure into residual blocks.

In HARU-Net, a Residual U-block is added after each down-sampling and up-sampling stage to extract local features and merge features from different levels. 

First, the input convolution is transformed into a feature map $F(x)$ with channels $C_{out}$. Then, $F(x)$ is fed into a symmetric encoder-decoder with a height of n, similar to U-Net. The value of n determines a larger receptive field and richer local and global features. Multi-scale features are extracted gradually from down-sampled feature maps and are encoded into a high-resolution feature map through step-by-step up-sampling, concatenation, and convolution. This process can alleviate the problem of losing fine details caused by direct upsampling with large ratios. Finally, the input feature $F(x)$ and the output feature $U(F(x))$ are fused by residual connection through summation.

It is noteworthy that the computational cost caused by the U-shaped structure is low because most operations are applied to the down-sampled feature maps. At the same time, the coefficient of the RSU on the quadratic term is much smaller, which brings excellent accuracy while improving efficiency.

\subsection{Hybrid Attention Module}

The Convolutional Block Attention Module (CBAM), proposed by Shanghyun Woo et al. \cite{b11}, is a module that combines spatial and channel attention mechanisms, addressing the limitations of single attention mechanisms (channel attention mechanism neglects spatial information interaction, while spatial attention mechanism neglects channel information interaction). The hybrid attention mechanism can fuse information from different levels of features to improve the network's understanding of different regions in input images. It consists of two key components: the spatial attention mechanism and the channel attention mechanism.

Channel attention module: the feature map is globally average-pooled and max-pooled separately to obtain two $1\times1\times C$ features, which are then passed through a multi-layer perceptron (MLP). The resulting features are concatenated and passed through a sigmoid activation function to obtain channel weight coefficients.

Spatial attention module: similar to the channel attention mechanism, the spatial attention mechanism performs average pooling and max pooling along the channel dimension to obtain two H×W×1 spatial features. The resulting features are concatenated and passed through a sigmoid activation function to obtain spatial weight coefficients.

The overall process of CBAM is shown in Fig. \ref{fig4}. The input feature map $F_{in}$ is passed through the channel attention mechanism $M_c$ to obtain a feature vector, which is multiplied with the input feature map $F_{in}$ to obtain $F_c$. $F_c$ is then passed through the spatial attention mechanism $M_s$ to obtain spatial weight $W_s$, which is multiplied with $F_c$ to obtain the final output feature $F_{out}$. The calculation formulas are as follows:

\begin{equation}
F_c = M_c(F_{in})\otimes F_{in}\label{eq1}
\end{equation}
\begin{equation}
F_{out} = M_s(F_{c})\otimes F_{c}\label{eq2}
\end{equation}

\subsection{Context Fusion Block}

In the context encoding layer of U2Net\cite{b10}, only concatenation is performed for each side branch without considering the importance of each branch. Inspired by SE-Net\cite{b35} and CE-Net\cite{b41}, we improved it by designing a CF-Block (context fusion) based on the SE-Block. This allows the model to capture different context information and improve the accuracy of semantic segmentation, as shown in Figure 1. The CF-Block helps the network focus on useful contextual information in the image, thereby enhancing segmentation accuracy and efficiency.
\begin{figure}[ht]
\centerline{\includegraphics[width=\columnwidth]{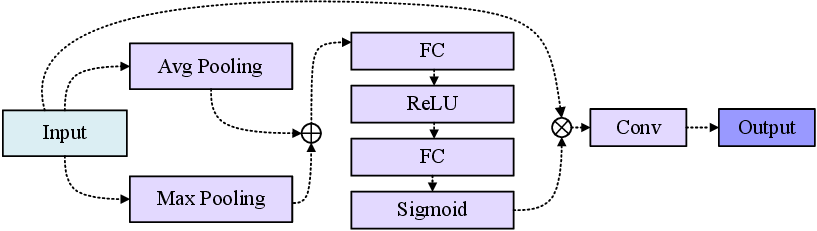}}
\caption{Structure of the Context Fusion Block is showed, where the input is the concatenation of context features from the backbone network (6 layers). The information is fused within the block and then outputted.}
\label{fig5}
\end{figure}

The process, as depicted in Fig. \ref{fig5}, involves linearly interpolating different-sized contexts to match the output size and then concatenating them as input to the CF-Block. Next, the feature maps are globally average-pooled and max-pooled separately to generate a feature vector. The feature vector is then subjected to a non-linear transformation to generate context feature layer weights. Finally, the input features are multiplied by the weights and upsampled to the size of the input image using a $3\times3$ convolutional layer.

\subsection{Architecture of HARU-Net}
HARU-Net is an instance segmentation model based on the U-Net architecture. As shown in Fig. \ref{fig2}, which can be divided into four main components: the encoder, decoder, attention module, and context mechanism.

The encoder stage consists of multiple downsampling layers, which perform convolution and pooling operations on the input image to extract features at different levels. The decoder stage also includes multiple upsampling layers, which upsample and merge the feature maps from the encoder stage to restore the original image resolution and output the segmentation predictions. Each downsampling and upsampling layer is composed of the Residual U-blocks mentioned earlier, and the nested U-structure enables the extraction of multi-scale features within stages and more effective aggregation of multi-level features across stages. The height of the Residual U-blocks is appropriately adjusted as the depth of the network increases.

Each upsampling layer consists of a Residual U-block and a skip connection, where the CBAM attention module extracts features from the downsampling information and combines them with the upsampling information through the skip connection to fuse low-level and high-level feature information, as shown in Fig. \ref{fig2}. The information extracted at a coarser scale is used for CBAM to eliminate irrelevant and noisy influences in the skip connection. This is performed before the concatenation operation to merge only relevant activations. Meanwhile, for deeper layers of the network, using a channel attention mechanism alone often proves to be more effective, and replacing CBAM with a channel attention mechanism at the deepest layer has yielded better results.

The last component is the feature fusion module, which generates probability maps. Similar to U2Net, our HARU-Net constructs two output branches, where each branch generates six probability maps, $\text{Mask}_i\text{; i}\in[1,6]$, from the last downsampling layer and all upsampling layers through a 3×3 convolutional layer and a sigmoid function. Subsequently, the feature maps from the side branches are upsampled to the size of the input image and fused using the CF-Block mentioned earlier to generate the final probability maps, $S_{mask}$ and $S_{edge}$.

HARU-Net aims to improve the network's ability to learn both fine details and global features in high-resolution medical images while keeping the computational complexity relatively low. This network model performs exceptionally well in nucleus instance segmentation tasks and accurately identifies segmentations even with limited samples.

\subsection{Supervision}
The output of the network model consists of two branches: foreground and contour. Each branch is composed of 6 probability maps. The loss function, denoted as $Loss$, is defined as follows:

\begin{equation}
\begin{split}
Loss = \omega_{edge}\tau_{edge}+\omega_{mask}\zeta_{mask}\\
+\sum_{i=1}^M(\omega_{side}^{(i)}\tau_{side}^{(i)}+\omega_{side}^{(i)}\zeta_{side}^{(i)})\label{eq3}
\end{split}
\end{equation}

Here, $\zeta_{side}^M$ and $\tau_{side}^M$ represent the losses for foreground and contour in the side branch, respectively. $\zeta_{mask}$ and $\tau_{mask}$ represent the losses for the final fused output feature maps $S_{mask}$ and $S_{edge}$. $\omega$ represents the weights for each loss term. For each term $\zeta$ and $\tau$, we calculate the loss as the sum of standard binary cross-entropy loss and Dice loss:

\begin{equation}
\begin{split}
    \text{CE}=-\sum_{(r,c)}^{(H,W)}[P_{G(r,c)}logPs(r,c) \\ 
    +(1-P_{G(r,c)})log(1-P_{s(r,c)})]
\end{split}
\end{equation}

\begin{equation}
    \text{Dice} = 1-\sum_{(r,c)}^{(H,W)}\begin{pmatrix} \frac{2\times P_{S(r,c)} \times P_{G(r,c)}}{ P_{S(r,c)}^2 +  P_{G(r,c)}^2}\end{pmatrix}
\end{equation}

\begin{equation}
    \zeta = \tau = CE+Dice
\end{equation}

$H$ and $W$ represent the height and width of the image, respectively. $(r,c)$ represents a pixel in the image, $P_{G(r,c)}$ represents the ground truth label for that pixel, and $P_{s(r,c)}$ represents the predicted probability of the model that the pixel is a positive label.

\subsection{Post Processing}
For nucleus segmentation, it is often necessary to segment instances. To address the issue of significant overlap between nuclei, DCAN\cite{b38} proposed a contour-based instance segmentation method. By subtracting the contours from the foreground, all overlapping nuclei are separated, resulting in individual cell instances. In deep learning models, if the annotated contours are very thin, the model training can easily get stuck in local minima, leading to poor segmentation results. By thickening the annotated contours, the model can escape local minima. However, this introduces a new problem. While subtracting the contours from the foreground allows obtaining instances of each cell, it also leads to the loss of foreground information.

\begin{figure}[ht]
\centerline{\includegraphics[width=\columnwidth]{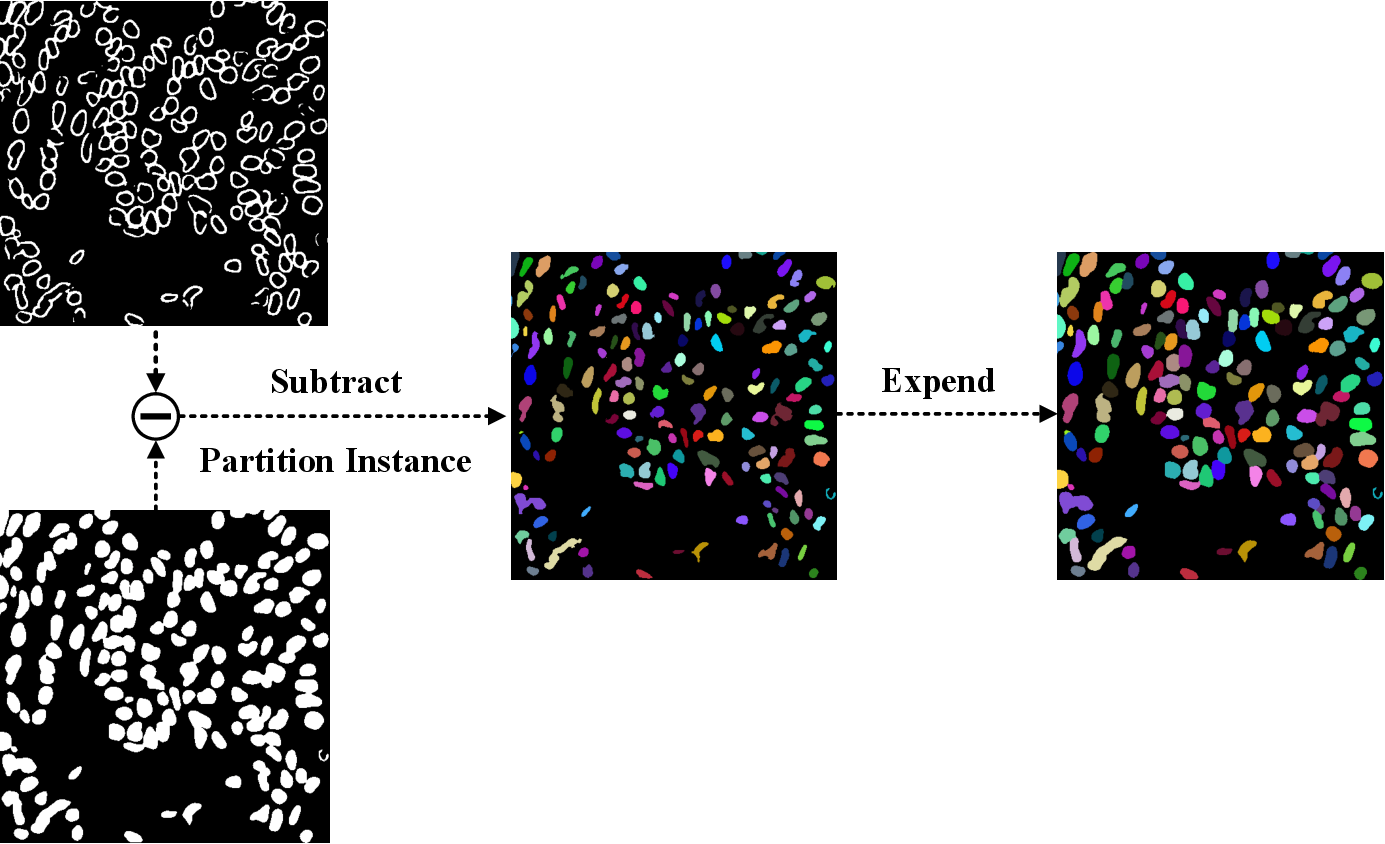}}
\caption{The input to the HARU-Net is the predicted semantic information and contour information. The semantic information is subtracted by the contour information to obtain instances. Then, each instance is repaired or refined.}
\label{fig6}
\end{figure}

\begin{algorithm}[ht]
    \renewcommand{\algorithmicrequire}{\textbf{Input:}}
    \renewcommand{\algorithmicensure}{\textbf{Output:}}
    \caption{Instance Segmentation Processing}
    \label{alg1}
    \begin{algorithmic}[1]
        \REQUIRE mask, edge
        
        \STATE $\text{kernel} \leftarrow \begin{bmatrix} 1 & 1 & 1 \\ 1 & 1 & 1 \\ 1 & 1 & 1 \end{bmatrix}$
        \STATE $\text{mask\_ins} \leftarrow \text{mask} \odot (1 - \text{edge})$
        \STATE $\text{num,objects,stats}  $\\ $\qquad\qquad\leftarrow\text{connected\_components\_with\_stats(mask\_ins)}$
        \WHILE{$\text{\textbf{not} is\_mask\_fully\_covered(objects, mask)}$}
        \FOR{$\text{mask\_id}=1$ \textbf{to} $\text{num\_objects}$}
        \STATE $\text{id\_mask} \leftarrow \text{objects[objects == mask\_id]}$
        
        \STATE $\text{id\_mask} \leftarrow \text{dilate\_mask(id\_mask, kernel)}$ 
        \STATE $\text{objects[(id\_mask != 0) \& (mask != 0) \& }$ \\ $\qquad\quad\;\text{(objects == 0)]}\leftarrow \text{mask\_id}$ 
        \ENDFOR
        \ENDWHILE
        
        \ENSURE $\text{objects}$
    \end{algorithmic}
\end{algorithm}

We have proposed a new method, as shown in Fig. \ref{fig6}. Here, we first subtract the contours from the foreground and perform instance labeling on each disconnected region. Then, we propagate the instance results outward until the foreground information is completely filled, generating the final segmentation mask. This method is simple and effective, as it can effectively utilize contour information to separate overlapping cells and restore cell instances using high-precision foreground information.

\section{Experiments and analysis}
In this section, we evaluate and test the performance and robustness of the proposed model through extensive experiments. This chapter includes the following contents: description of four publicly available datasets (MoNuSeg, CoNSep, CPM-17, BNS), dataset preprocessing, evaluation metrics, ablation experiments, and comparative experiments.

\subsection{Dataset}
To demonstrate the effectiveness of the algorithm, we selected four publicly available datasets for nuclear segmentation and conducted a comparative analysis. The descriptions of the datasets are as follows:

\textbf{MoNuSeg}\cite{MoNuSeg} dataset include  a training set with 30 images from seven organs with annotations of 21,623 individual nuclei. A test dataset with 14 images taken from seven organs, including two organs that did not appear in the training set was released without annotations. All images are in color and the size is $1000\times1000$.

\textbf{CoNSep}\cite{b12} dataset consists of 41 H\&E stained images of colorectal adenocarcinoma (CRA) with a size of $1000\times 1000$ pixels. This dataset was collected by University Hospitals Coventry and Warwickshire using a $40\times$ scanner. We employed the same dataset as mentioned in the reference for segmentation. The training set comprises 27 images, while the testing set consists of 14 images.

\textbf{CPM-17}\cite{cpm} dataset consists of 64 H\&E stained pathology images with pixel-level annotations. The training set contains 32 images, and the test set contains 32 images. Each image was scanned at 40x magnification and has a size ranging from $500\times500$ to $600\times600$.

\textbf{BNS(TNBC)}\cite{BNS} dataset released by the Curie Institute, focuses on "Triple-Negative Breast Cancer" (TNBC). This dataset includes 50 histopathological images stained with H\&E at a magnification of $40\times$. The dataset provides annotations for 4,056 nucleus boundaries. Each image has a size of $512\times 512$ pixels. For our experiments, we randomly selected 20 images as the test set and used the remaining 30 images as the training set.

\subsection{Evaluation metrics}

In this experiment, we employed Dice, Panoptic Quality (PQ), and Aggregated Jaccard Index (AJI) as performance evaluation metrics. These metrics are defined as follows:

\textbf{Dice}: The Dice measures the similarity at the pixel level and is computed using the formula:

\begin{equation}
    Dice=\frac{2\times TP}{2\times TP+FP+FN}
\end{equation}

TP (True Positive) represents the accurately segmented regions, FP (False Positive) represents the regions erroneously segmented as positive examples, and FN (False Negative) represents the regions erroneously segmented as negative examples.

\textbf{Panoptic Quality (PQ)}  is used to evaluate the performance of instance segmentation algorithms by assessing the matching between predicted results and ground truth labels. Its formula is defined as follows:

\begin{equation}
    PQ=\frac{2\times TP}{2\times TP+FP+FN}\times\frac{\sum_{(x,y)\in TP}IoU(x,y)}{|TP|}
\end{equation}

The first part corresponds to the Precision-Recall component, which measures the accuracy of the segmentati    on results. TP, FP, and FN have the same definitions as in the Dice. The second part represents the IoU component, which evaluates the precision of the segmentation results. The term $\sum_{(x,y)\in TP}IoU(x,y)$ denotes the average IoU value of correctly matched objects, and TP indicates the number of correctly matched objects.

\textbf{Aggregated Jaccard Index (AJI)} is utilized to assess the similarity between the instance segmentation results generated by an algorithm and the ground truth segmentations. Its formula is defined as follows:
\begin{equation}
    AJI=\frac{\sum^N_{i=1}|G_i\cap P_M^i|}{\sum_{i=1}^N|G_i\cup P_M^i|+\sum_{F\in U}|P_F|}
\end{equation}

In the formula for AJI, $N$ represents the number of instances in the sample. $G_i$ denotes the ground truth mask of the $i$-th instance, and $P_M^i$ represents the predicted mask of the $i$-th instance. The term $\sum_{i=1}^N|G_i\cup P_M^i|$ denotes the total size of the union between the matched predicted and ground truth segmentations. The second term $\sum_{F\in U}|P_F|$ represents the total number of pixels in unmatched predicted segmentations. The numerator of the formula, $\sum^N_{i=1}|G_i\cap P_M^i|$, indicates the total number of pixels that are correctly matched.

\subsection{Implementation details}
We implemented the models using PyTorch version 1.18.0 on a system equipped with an NVIDIA GeForce RTX 3090 GPU. During the training process, we retained the original sizes of all datasets and applied data augmentation techniques, including random horizontal and vertical flips and random rotations within a specified range. We employed Stochastic Gradient Descent (SGD) as the optimization method with a learning rate of 1e-5 and an adaptive learning rate decay strategy. The input sizes are maintained as follows: MoNuSeg and CoNSep had an input size of $1000\times1000$ with a batch size of 4, cpm-17 had an input size of $512\times512$ with a batch size of 12, and BNS had an input size of $600\times600$. For BNS, we randomly selected 15 images for testing and used the remaining 25 images for training, with a batch size of 12. The models are trained for 200 iterations, and checkpoints are saved at the end of each epoch. For the CoNSep dataset, characterized by high levels of noise, significant boundary blurring, and diverse cell shapes, we employed additional data augmentation techniques such as feature enhancement, image denoising, and filtering. Moreover, in the post-processing step, we employed specific methods to address the segmentation of overlapping nuclei.

\subsection{Comparison with other methods}

To further evaluate HARU-Net, we conducted comparative experiments on four publicly available datasets: MoNuSeg, CoNSep, BNS, and CPM-17. We compared HARU-Net with state-of-the-art models as well as classical models to assess its performance.
\subsubsection{Evaluation on MoNuSeg Dataset}
\ 
\newline
\indent The MoNuSeg dataset is a medical imaging dataset released by Kumar et al. in 2017, designed to support research on tumor segmentation tasks. The training dataset consists of 30 images, and the test dataset contains 14 images. The images have a size of $1000\times1000$ pixels. To augment the training dataset, we applied techniques such as cropping and flipping, resulting in a total of 240 augmented training images. The diversity of this dataset comes from the variations in the appearance of nuclei from different organs and patients, as well as the staining protocols used by multiple hospitals. This diversity helps in the development of robust and generalizable nucleus segmentation techniques.

\begin{table}[ht]
\centering
\caption{Comparison with other models on the MoNuSeg}
\label{table2}
\begin{tabular}{c|ccc}
\hline
\multirow{2}{*}{\textbf{Model}} & \multicolumn{3}{c}{\textbf{MoNuSeg}} \\ \cline{2-4} 
                                & \textbf{Dice} & \textbf{AJI} & \textbf{PQ}      \\ \hline
CNN\cite{b34}                             & 0.762   & 0.508   &         \\
U-Net\cite{b29}                           & 0.758   & 0.556   & 0.478   \\
Mask R-CNN\cite{b9}                      & 0.760   & 0.546   & 0.509   \\
DITS\cite{b48}                            & 0.786   & 0.560   & 0.443   \\
DCAN\cite{b38}                            & 0.793   & 0.525   & 0.492   \\
Micro-Net\cite{b49}                       & 0.797   & 0.560   & 0.519   \\
HoVer-Net\cite{b12}                       & 0.826   & 0.618   & 0.597   \\
AL-Net\cite{b33}                          & 0.823   & 0.649   & 0.610   \\
\textbf{HARU-Net}                       & \textbf{0.838}   & \textbf{0.656}   & \textbf{0.624}   \\ \hline
\end{tabular}
\end{table}

The evaluation results of all the networks on the MoNuSeg dataset are presented in Table \ref{table2}. We have listed the results of other networks as reported in their respective papers. It can be observed that HARU-Net achieves significantly higher Dice, AJI, and PQ scores compared to all other networks, with improvements of 1.3\%, 0.7\%, and 1.4\%, respecti    vely. Furthermore, the performance of HARU-Net on this dataset demonstrates its superiority over other networks without excessive data augmentation.

\subsubsection{Evaluation on CoNSep Dataset}
\ 
\newline
\indent The CoNSep dataset consists of 16 Whole Slide Images (WSI) of colorectal adenocarcinoma (CRA), where each WSI belongs to a separate patient and was scanned using the Omnyx VL120 scanner at the Pathology Department of the University Hospitals Coventry and Warwickshire NHS Trust in the United Kingdom. The nucleus morphology in CoNSep exhibits noticeable deviations from general cancer datasets, and both the training and testing sets contain a significant amount of noise and blurry boundaries. To address these challenges, we have made further improvements to the dataset. In the post-processing stage, we introduced an erosion operation, which has led to an improvement in the instance segmentation accuracy of the model.

\begin{table}[ht]
\centering
\caption{Comparison with other models on the CoNSep}
\label{table4}
\begin{tabular}{c|ccc}
\hline
\multirow{2}{*}{\textbf{Model}} & \multicolumn{3}{c}{\textbf{CoNSep}}        \\ \cline{2-4} 
                                & \textbf{Dice} & \textbf{AJI} & \textbf{PQ} \\ \hline
U-Net\cite{b29}                          & 0.724         & 0.482        & 0.328       \\
Mask R-CNN\cite{b9}                      & 0.740         & 0.474        & 0.460       \\
DIST\cite{b48}                           & 0.804         & 0.502        & 0.398       \\
DCAN\cite{b38}                           & 0.733         & 0.289        & 0.256       \\
HoVer-Net\cite{b12}                      & 0.853         & \textbf{0.571}        & \textbf{0.547}       \\
\textbf{HARU-Net}                        & \textbf{0.856}         & 0.554        & 0.518       \\ \hline
\end{tabular}
\end{table}
The evaluation results of all networks on the HARU-Net dataset are shown in Table 
\ref{table4}. The Dice, AJI, and PQ scores of HARU-Net are 85.6, 55.4, and 51.8, respectively. It can be observed that HARU-Net still achieves a high Dice score, but the AJI and PQ scores are relatively lower. Upon careful investigation, it was found that the blurred boundaries of overlapping regions pose challenges in the post-processing stage, making it difficult to completely separate overlapping cells. Although the erosion operation helps to improve the distinction of overlapping cells, further improvements are still needed to enhance the overall performance. HoVer-Net model achieves superior instance segmentation accuracy by leveraging distance-based predictions, effectively addressing the issue of blurred nucleus boundaries. As a result, HoVer-Net outperforms other models in terms of instance segmentation precision.

\subsubsection{Evaluation on BNS Dataset}
\ 
\newline
\indent The BNS dataset consists of 50 images obtained from a study on triple-negative breast cancer. Each tissue slice was scanned to obtain Whole Slide Images (WSI) with a size of $512\times 512$. The dataset includes different types of cells, such as normal epithelial cells, myoepithelial breast cells, invasive cancer cells, and fibroblast cells. There are a total of 2754 nuclei in the entire dataset, with the number of nuclei per image ranging from 5 to 293. Out of the 50 images, 35 images were selected as the training set, and 15 images were used as the test set. Data augmentation techniques such as flipping and rotation were applied to the training images, resulting in a total of 280 augmented training images. The test set consists of 120 images.
\begin{table}[ht]
\centering
\caption{Comparison with other models on the BNS}
\label{table3}
\begin{tabular}{c|ccc}
\hline
\multirow{2}{*}{\textbf{Model}} & \multicolumn{3}{c}{\textbf{BNS}}           \\ \cline{2-4} 
                                & \textbf{Dice} & \textbf{AJI} & \textbf{PQ} \\ \hline
U-Net++\cite{b50}                         & 0.777         & 0.572        & 0.549       \\
Attention U-Net\cite{b30}                 & 0.782         & 0.576        & 0.554       \\
CE-Net\cite{b31}                          & 0.779         & 0.582        & 0.548       \\
Joint segmentation\cite{b51}              & 0.781         & 0.587        & 0.549       \\
AL-Net\cite{b33}                          & 0.790         & 0.624        & 0.617       \\
\textbf{HARU-Net}                        & \textbf{0.812}         & \textbf{0.651}        & \textbf{0.631}       \\ \hline
\end{tabular}
\end{table}

The evaluation results of all the networks on the BNS dataset are presented in Table \ref{table3}. HARU-Net achieves Dice, AJI, and PQ scores of 81.2, 65.1, and 63.1, respectively. Due to the presence of attention mechanisms, HARU-Net demonstrates exceptional performance on this dataset, while other networks may suffer from issues such as over-segmentation and under-segmentation.
\subsubsection{Evaluation on CPM-17 Dataset}
\ 
\newline
\indent CPM-17 dataset comprises tissue images from patients with non-small cell lung cancer (NSCLC), head and neck squamous cell carcinoma (HNSCC), glioblastoma multiforme (GBM), and low-grade glioma (LGG). For this dataset, we applied only flip and translation data augmentation techniques while maintaining the original image size for input. HARU-Net achieved Dice, AJI, and PQ scores of 8.95, 72.1, and 70.1 in Table \ref{table5}, respectively. It is evident that HARU-Net performs remarkably well on this dataset, showing an improvement of around 2\% to 3\% in all metrics.
\begin{table}[ht]
\centering
\caption{Comparison with other models on the cpm-17}
\label{table5}
\begin{tabular}{c|ccc}
\hline
\multirow{2}{*}{\textbf{Model}} & \multicolumn{3}{c}{\textbf{CPM-17}}        \\ \cline{2-4} 
                                & \textbf{Dice} & \textbf{AJI} & \textbf{PQ} \\ \hline
U-Net\cite{b29}                           & 0.813         & 0.643        & 0.578       \\
Mask R-CNN\cite{b9}                      & 0.850         & 0.684        & 0.674       \\
DIST\cite{b48}                            & 0.826         & 0.616        & 0.504       \\
DCAN\cite{b38}                            & 0.828         & 0.561        & 0.545       \\
HoVer-Net\cite{b12}                       & 0.869         & 0.705        & 0.697       \\
\textbf{HARU-Net}                        & \textbf{0.895}         & \textbf{0.721}        & \textbf{0.701}       \\ \hline
\end{tabular}
\end{table}

\subsection{Ablation study}

To evaluate each module of HARU-Net, we conducted a series of ablation experiments on the MoNuSeg dataset. In these experiments, we chose U-Net as the baseline model and used watershed as the instance segmentation method for all experiments except the last one.

\begin{table}[ht]

\centering
\caption{ Qualitative results of ablation studies. Impact of different modules on network performance.}
\label{table1}
\begin{tabular}{llll}
\hline
\multicolumn{1}{c}{\textbf{Model}}                                                         & \textbf{Dice}  & \textbf{AJI}   & \textbf{PQ}    \\ \hline
Baseline U-Net                                                                             & 0.758          & 0.478          & 0.556          \\
U-Net + RSU(U2Net)                                                                         & 0.790          & 0.533          & 0.570          \\
U-Net + RSU + CF-Block                                                                     & 0.808          & 0.580          & 0.579          \\
U-Net + RSU + CBAM                                                                         & 0.820          & 0.590          & 0.583          \\
\begin{tabular}[c]{@{}l@{}}U-Net + RSU + CBAM\\ + CF-Block\end{tabular}                    & 0.827          & 0.598          & 0.589          \\
\begin{tabular}[c]{@{}l@{}}U-Net + RSU + CBAM \\ + CF-Block + Post Processing\end{tabular} & \textbf{0.838} & \textbf{0.656} & \textbf{0.624} \\ \hline
\end{tabular}
\end{table}
 
\subsubsection{Ablation on Residual U-blocks(RSU)}

By incorporating the RSU module into the main task input of the base model, we observed a significant improvement in fine-grained object extraction. The Base + RSU model achieved a 3.0\% increase in AJI, a 1.4\% increase in the Dice score, and a 5.5\% increase in the PQ score. The Residual U-Block effectively extracts multi-scale features within a single stage, leading to noticeable improvements in the model's performance.

\subsubsection{Ablation on Convolutional Block Attention Module(CBAM)}
In this experiment, we incorporated the CBAM module into the U2Net for the task of nuclear segmentation. Significant improvements are achieved, where the U-Net + RSU + CBAM model showed increases of 2.2\% in Dice score, 1.3\% in PQ score, and 6.7\% in AJI score. Furthermore, to demonstrate the effectiveness of using channel attention mechanism for the deepest level attention module, we compared two methods: one with CBAM attention mechanism applied at all levels and the other with the channel attention mechanism applied only at the deepest level. Both methods showed improvements of 0.6\% in both Dice score and PQ score. This is because the deepest level of the network has a small spatial size (32x32), and applying a spatial attention mechanism would have a negative impact on the overall model performance.
\subsubsection{Ablation on Context Fusion Block(CF-Block)}
In this experiment, we validated the effectiveness of the CF-Block in the model. The CF-Block allows weighted fusion of contextual information instead of simple concatenation, leading to superior performance. The U-Net + RSU + CBAM + CF-Block model showed improvements of 0.7\%, 0.6\%, and 0.9\% in the three evaluation metrics compared to the U-Net + RSU + CBAM model.
\subsubsection{Ablation on Post Processing}
In this experiment, we validated the effectiveness of Post Processing in instance segmentation. The three evaluation metrics of the model showed significant improvements of 1.0\%, 3.5\%, and 5.8\%, respectively. Additionally, we discovered that the auxiliary task of edge production also contributed to improved pixel prediction for the nucleus in the main task. By combining contour and semantic analysis, the model not only better-distinguished nucleus instances but also accurately restored foreground information for each instance, resulting in superior performance.

\section{Discussion}
Pathology diagnosis plays a crucial role in medical diagnosis, especially in cancer diagnosis. To improve the efficiency and robustness of automated histopathology image analysis, we propose a hybrid attention dual-branch network based on Residual U-blocks. Extensive experiments are conducted on four challenging datasets to demonstrate the effectiveness of the network model and its individual modules. In the multi-task network, contour extraction and object segmentation are two complementary tasks. Previous works that divide instances based on contour extraction, such as DCAN and AL-Net, only consider nucleus instance division without focusing on the segmentation accuracy of each instance. Therefore, we propose to propagate around each nucleus instance until the target segmentation result is completely filled. Experimental results show that this method is simple yet effective.

From the MoNuSeg, CPM-17, and BNS datasets, we observe that the model achieves high segmentation accuracy and can effectively segment overlapping nuclei. However, in the CoNSep dataset, the instance segmentation accuracy is not satisfactory. Through careful investigation, we found that the shape prior of nuclei in the CoNSep dataset deviates significantly from the general cancer dataset. The nuclei boundaries in this dataset are more blurred, leading to unsatisfactory contour extraction results for overlapping regions. Therefore, for this dataset, we applied an erosion operation in the post-processing stage to improve the model's accuracy. After in-depth analysis, the following improvements can be made in future work:
\begin{itemize}
\item Address the issue of unclear contours and incomplete extraction of overlapping region contours. Consider incorporating prior knowledge of nucleus morphological features into the loss function to further optimize contour extraction accuracy or explore specific techniques for contour extraction in overlapping regions.
\item HARU-Net demonstrates good performance in semantic segmentation and is not limited to cervical cell classification, indicating its potential application in other medical image classification tasks.
\item In the post-processing stage, the instance spreading technique for each nucleus does not accurately handle overlapping boundaries. Further improvements can be made to refine the spreading strategy.
\end{itemize}

\section{Conclusion}

This study addresses the complex problem of nucleus segmentation and proposes a hybrid attention dual-branch network based on Residual U-blocks. The network simultaneously predicts the semantic and contour information of nuclei and achieves high-precision instance segmentation through post-processing. Additionally, we discovered that the backpropagation from the auxiliary task further improves the accuracy of the main task. The key components of the network include Residual U-blocks, hybrid attention mechanism, and context encoding layers. The network model achieves highly accurate semantic segmentation. Moreover, our post-processing method demonstrates strong generality and effectiveness by leveraging multi-task information (semantic and contour information) to delineate object instances. In the experiments, we validate the critical roles of different components in the segmentation task model. Extensive experimental results on four challenging histopathology image segmentation tasks demonstrate the superiority of our method, surpassing state-of-the-art approaches.






\end{sloppypar}

\begin{thebibliography}{00}
\bibitem{b1}
F.~Clayton, ``Pathologic correlates of survival in 378 lymph node-negative
  infiltrating ductal breast carcinomas. mitotic count is the best single
  predictor,'' \emph{Ca Cancer J. Clin.}, vol.~68, no.~6, pp. 1309--1317, 1991.

\bibitem{b2}
C.~W. Elston and I.~O. Ellis, ``Pathological prognostic factors in breast
  cancer. i. the value of histological grade in breast cancer: experience from
  a large study with long-term follow-up,'' \emph{Histopathology.}, vol.~19,
  no.~5, pp. 403--410, 1991.

\bibitem{b3}
H.~Sung, J.~Ferlay, R.~L. Siegel, M.~Laversanne, I.~Soerjomataram, A.~Jemal,
  and F.~Bray, ``Global cancer statistics 2020: Globocan estimates of incidence
  and mortality worldwide for 36 cancers in 185 countries,'' \emph{CA Cancer J
  Clin.}, vol.~71, no.~3, pp. 209--249, 2021.

\bibitem{b4}
E.~Davey, A.~Barratt, L.~Irwig, S.~F. Chan, P.~Macaskill, P.~Mannes, and A.~M.
  Saville, ``Effect of study design and quality on unsatisfactory rates,
  cytology classifications, and accuracy in liquid-based versus conventional
  cervical cytology: a systematic review,'' \emph{Lancet.}, vol. 367, no. 9505,
  pp. 122--132, 2006.

\bibitem{b5}
L.~Zhang, H.~Kong, C.~Ting~Chin, S.~Liu, X.~Fan, T.~Wang, and S.~Chen,
  ``Automation-assisted cervical cancer screening in manual liquid-based
  cytology with hematoxylin and eosin staining,'' \emph{Cytometry A.}, vol.~85,
  no.~3, pp. 214--230, 2014.

\bibitem{b6}
O.~Sertel, J.~Kong, H.~Shimada, U.~Catalyurek, J.~H. Saltz, and M.~Gurcan,
  ``Computer-aided prognosis of neuroblastoma: classification of stromal
  development on whole-slide images,'' in \emph{Med. Imaging 2008:
  Comput-Aided. Diagnosis}, vol. 6915.\hskip 1em plus 0.5em minus 0.4em\relax
  SPIE, 2008, pp. 211--220.

\bibitem{b7}
K.-H. Chow, R.~E. Factor, and K.~S. Ullman, ``The nuclear envelope environment
  and its cancer connections,'' \emph{Nature Reviews Cancer}, vol.~12, no.~3,
  pp. 196--209, 2012.

\bibitem{a1}
H.~He, Z.~Huang, Y.~Ding, G.~Song, L.~Wang, Q.~Ren, P.~Wei, Z.~Gao, and
  J.~Chen, ``Cdnet: Centripetal direction network for nuclear instance
  segmentation,'' in \emph{Proc. IEEE Int. Conf. Comput. Vision. (ICCV)}, 2021,
  pp. 4026--4035.

\bibitem{a2}
N.~Alemi~Koohbanani, M.~Jahanifar, A.~Gooya, and N.~Rajpoot, ``Nuclear instance
  segmentation using a proposal-free spatially aware deep learning framework,''
  in \emph{Med. Image Comput. Comput. Assist. Interv. (MICCAI)}.\hskip 1em plus
  0.5em minus 0.4em\relax Springer, 2019, pp. 622--630.

\bibitem{stardist}
U.~Schmidt, M.~Weigert, C.~Broaddus, and G.~Myers, ``Cell detection with
  star-convex polygons,'' in \emph{Med. Image Comput. Comput. Assist. Interv.
  (MICCAI)}.\hskip 1em plus 0.5em minus 0.4em\relax Springer, 2018, pp.
  265--273.

\bibitem{cppnet}
S.~Chen, C.~Ding, M.~Liu, J.~Cheng, and D.~Tao, ``Cpp-net: Context-aware
  polygon proposal network for nucleus segmentation,'' \emph{IEEE Trans. Image
  Process.}, 2023.

\bibitem{a3}
S.~Wienert, D.~Heim, K.~Saeger, A.~Stenzinger, M.~Beil, P.~Hufnagl, M.~Dietel,
  C.~Denkert, and F.~Klauschen, ``Detection and segmentation of cell nuclei in
  virtual microscopy images: a minimum-model approach,'' \emph{Sci. Rep.},
  vol.~2, no.~1, p. 503, 2012.

\bibitem{a4}
C.~Chen, W.~Wang, J.~A. Ozolek, and G.~K. Rohde, ``A flexible and robust
  approach for segmenting cell nuclei from 2d microscopy images using
  supervised learning and template matching,'' \emph{Cytometry A.}, vol.~83,
  no.~5, pp. 495--507, 2013.

\bibitem{b8}
B.~E. Bejnordi, G.~Litjens, N.~Timofeeva, I.~Otte-H{\"o}ller, A.~Homeyer,
  N.~Karssemeijer, and J.~A. Van Der~Laak, ``Stain specific standardization of
  whole-slide histopathological images,'' \emph{IEEE Trans. Image Process.},
  vol.~35, no.~2, pp. 404--415, 2015.

\bibitem{b9}
K.~He, G.~Gkioxari, P.~Doll{\'a}r, and R.~Girshick, ``Mask r-cnn,'' in
  \emph{Proc. IEEE Int. Conf. Comput. Vision. (ICCV)}, 2017, pp. 2961--2969.

\bibitem{b10}
X.~Qin, Z.~Zhang, C.~Huang, M.~Dehghan, O.~R. Zaiane, and M.~Jagersand,
  ``U2-net: Going deeper with nested u-structure for salient object
  detection,'' \emph{Pattern Recognit.}, vol. 106, p. 107404, 2020.

\bibitem{b11}
S.~Woo, J.~Park, J.-Y. Lee, and I.~S. Kweon, ``Cbam: Convolutional block
  attention module,'' in \emph{Proc. Eur. Conf. Comput. Vis. (ECCV)}, 2018, pp.
  3--19.

\bibitem{b12}
S.~Graham, Q.~D. Vu, S.~E.~A. Raza, A.~Azam, Y.~W. Tsang, J.~T. Kwak, and
  N.~Rajpoot, ``Hover-net: Simultaneous segmentation and classification of
  nuclei in multi-tissue histology images,'' \emph{Med. Image Anal.}, vol.~58,
  p. 101563, 2019.

\bibitem{b13}
T.~Ilyas, Z.~I. Mannan, A.~Khan, S.~Azam, H.~Kim, and F.~De~Boer, ``Tsfd-net:
  Tissue specific feature distillation network for nuclei segmentation and
  classification,'' \emph{Neural Netw.}, vol. 151, pp. 1--15, 2022.

\bibitem{b14}
Q.~D. Vu, S.~Graham, T.~Kurc, M.~N.~N. To, M.~Shaban, T.~Qaiser, N.~A.
  Koohbanani, S.~A. Khurram, J.~Kalpathy-Cramer, T.~Zhao \emph{et~al.},
  ``Methods for segmentation and classification of digital microscopy tissue
  images,'' \emph{Front. Bioeng. Biotechnol.}, p.~53, 2019.

\bibitem{MoNuSeg}
N.~Kumar, R.~Verma, D.~Anand, Y.~Zhou, O.~F. Onder, E.~Tsougenis, H.~Chen,
  P.-A. Heng, J.~Li, Z.~Hu \emph{et~al.}, ``A multi-organ nucleus segmentation
  challenge,'' \emph{{IEEE} Trans. Med. Imag.}, vol.~39, no.~5, pp. 1380--1391,
  2019.

\bibitem{BNS}
P.~Naylor, M.~La{\'e}, F.~Reyal, and T.~Walter, ``Nuclei segmentation in
  histopathology images using deep neural networks,'' in \emph{IEEE Comput.
  Soc. Conf. Comput. Vis. Pattern Recogn.}\hskip 1em plus 0.5em minus
  0.4em\relax IEEE, 2017, pp. 933--936.

\bibitem{cpm}
Q.~D. Vu, S.~Graham, T.~Kurc, M.~N.~N. To, M.~Shaban, T.~Qaiser, N.~A.
  Koohbanani, S.~A. Khurram, J.~Kalpathy-Cramer, T.~Zhao \emph{et~al.},
  ``Methods for segmentation and classification of digital microscopy tissue
  images,'' \emph{Front. Bioeng. Biotechnol.}, p.~53, 2019.

\bibitem{b18}
F.~Xing and L.~Yang, ``Robust nucleus/cell detection and segmentation in
  digital pathology and microscopy images: a comprehensive review,''
  \emph{{IEEE} Rev. Biomed. Eng.}, vol.~9, pp. 234--263, 2016.

\bibitem{b19}
J.~C. Caicedo, J.~Roth, A.~Goodman, T.~Becker, K.~W. Karhohs, M.~Broisin,
  C.~Molnar, C.~McQuin, S.~Singh, F.~J. Theis \emph{et~al.}, ``Evaluation of
  deep learning strategies for nucleus segmentation in fluorescence images,''
  \emph{Cytometry A.}, vol.~95, no.~9, pp. 952--965, 2019.

\bibitem{b20}
C.~W{\"a}hlby, I.-M. Sintorn, F.~Erlandsson, G.~Borgefors, and E.~Bengtsson,
  ``Combining intensity, edge and shape information for 2d and 3d segmentation
  of cell nuclei in tissue sections,'' \emph{J. Microsc.}, vol. 215, no.~1, pp.
  67--76, 2004.

\bibitem{b21}
Y.~Y. Boykov and M.-P. Jolly, ``Interactive graph cuts for optimal boundary \&
  region segmentation of objects in nd images,'' in \emph{Proc. IEEE Int. Conf.
  Comput. Vision. (ICCV)}, vol.~1.\hskip 1em plus 0.5em minus 0.4em\relax IEEE,
  2001, pp. 105--112.

\bibitem{b22}
D.~Stutz, A.~Hermans, and B.~Leibe, ``Superpixels: An evaluation of the
  state-of-the-art,'' \emph{Comput. Vision Image Understanding}, vol. 166, pp.
  1--27, 2018.

\bibitem{b23}
X.~Bai, C.~Sun, and C.~Sun, ``Cell segmentation based on fopso combined with
  shape information improved intuitionistic fcm,'' \emph{{IEEE} J. Biomed.
  Health Inform.}, vol.~23, no.~1, pp. 449--459, 2018.

\bibitem{b24}
C.-H. Lin and C.-C. Chen, ``Image segmentation based on edge detection and
  region growing for thinprep-cervical smear,'' \emph{Int. J. Pattern Recognit.
  Artif. Intell.}, vol.~24, no.~07, pp. 1061--1089, 2010.

\bibitem{b25}
Y.~Song, W.~Cai, D.~D. Feng, and M.~Chen, ``Cell nuclei segmentation in
  fluorescence microscopy images using inter-and intra-region discriminative
  information,'' in \emph{Proc. Annu. Int. Conf. IEEE Eng. Med. Biol. Soc.
  (EMBC)}.\hskip 1em plus 0.5em minus 0.4em\relax IEEE, 2013, pp. 6087--6090.

\bibitem{b26}
F.~Xing, Y.~Xie, and L.~Yang, ``An automatic learning-based framework for
  robust nucleus segmentation,'' \emph{{IEEE} Trans. Med. Imag.}, vol.~35,
  no.~2, pp. 550--566, 2015.

\bibitem{b27}
G.~Lv, K.~Wen, Z.~Wu, X.~Jin, H.~An, and J.~He, ``Nuclei r-cnn: Improve mask
  r-cnn for nuclei segmentation,'' in \emph{IEEE Int. Conf. Inf. Commun. Signal
  Process. (ICICSP)}.\hskip 1em plus 0.5em minus 0.4em\relax IEEE, 2019, pp.
  357--362.

\bibitem{b28}
J.~Dai, Y.~Li, K.~He, and J.~Sun, ``R-fcn: Object detection via region-based
  fully convolutional networks,'' \emph{Adv. neural inf. proces. syst.},
  vol.~29, 2016.

\bibitem{b29}
O.~Ronneberger, P.~Fischer, and T.~Brox, ``U-net: Convolutional networks for
  biomedical image segmentation,'' in \emph{Med. Image Comput. Comput. Assist.
  Interv. (MICCAI)}.\hskip 1em plus 0.5em minus 0.4em\relax Springer, 2015, pp.
  234--241.

\bibitem{b30}
O.~Oktay, J.~Schlemper, L.~L. Folgoc, M.~Lee, M.~Heinrich, K.~Misawa, K.~Mori,
  S.~McDonagh, N.~Y. Hammerla, B.~Kainz \emph{et~al.}, ``Attention u-net:
  Learning where to look for the pancreas,'' \emph{Proc. IEEE Conf. Comput. Vis. Pattern Recognit. (CVPR)}, 2018.

\bibitem{b31}
Z.~Gu, J.~Cheng, H.~Fu, K.~Zhou, H.~Hao, Y.~Zhao, T.~Zhang, S.~Gao, and J.~Liu,
  ``Ce-net: Context encoder network for 2d medical image segmentation,''
  \emph{{IEEE} Trans. Med. Imag.}, vol.~38, no.~10, pp. 2281--2292, 2019.

\bibitem{b32}
Y.~Zhou, O.~F. Onder, Q.~Dou, E.~Tsougenis, H.~Chen, and P.-A. Heng, ``Cia-net:
  Robust nuclei instance segmentation with contour-aware information
  aggregation,'' in \emph{Lect. Notes Comput. Sci.}\hskip 1em plus 0.5em minus
  0.4em\relax Springer, 2019, pp. 682--693.

\bibitem{b33}
J.~Zhao, Y.-J. He, S.-Q. Zhao, J.-J. Huang, and W.-M. Zuo, ``Al-net: Attention
  learning network based on multi-task learning for cervical nucleus
  segmentation,'' \emph{{IEEE} J. Biomed. Health Inform.}, vol.~26, no.~6, pp.
  2693--2702, 2021.

\bibitem{b34}
F.~Wang, M.~Jiang, C.~Qian, S.~Yang, C.~Li, H.~Zhang, X.~Wang, and X.~Tang,
  ``Residual attention network for image classification,'' in \emph{Proc. Eur. Conf. Comput. Vis. (ECCV)}, 2017, pp. 3156--3164.

\bibitem{b35}
J.~Hu, L.~Shen, and G.~Sun, ``Squeeze-and-excitation networks,'' in \emph{Proc.
  IEEE Conf. Comput. Vis. Pattern Recognit.}, 2018, pp. 7132--7141.

\bibitem{b36}
L.~Chen, H.~Zhang, J.~Xiao, L.~Nie, J.~Shao, W.~Liu, and T.-S. Chua, ``Sca-cnn:
  Spatial and channel-wise attention in convolutional networks for image
  captioning,'' in \emph{Proc. IEEE Conf. Comput. Vis. Pattern Recognit.},
  2017, pp. 5659--5667.

\bibitem{b37}
S.~Woo, J.~Park, J.-Y. Lee, and I.~S. Kweon, ``Cbam: Convolutional block
  attention module,'' in \emph{Proc. Eur. Conf. Comput. Vis. (ECCV)}, 2018, pp.
  3--19.

\bibitem{b38}
H.~Chen, X.~Qi, L.~Yu, Q.~Dou, J.~Qin, and P.-A. Heng, ``Dcan: Deep
  contour-aware networks for object instance segmentation from histology
  images,'' \emph{Med. Image Anal.}, vol.~36, pp. 135--146, 2017.

\bibitem{b39}
L.~Sun, S.~Cheng, Y.~Zheng, Z.~Wu, and J.~Zhang, ``Spanet: Successive pooling
  attention network for semantic segmentation of remote sensing images,''
  \emph{{IEEE} J. Sel. Topics Appl. Earth Observ. Remote Sens.}, vol.~15, pp.
  4045--4057, 2022.

\bibitem{b40}
S.~Chen, C.~Ding, and D.~Tao, ``Boundary-assisted region proposal networks for
  nucleus segmentation,'' in \emph{Med. Image Comput. Comput. Assist. Interv.
  (MICCAI)}.\hskip 1em plus 0.5em minus 0.4em\relax Springer, 2020, pp.
  279--288.

\bibitem{b41}
Z.~Gu, J.~Cheng, H.~Fu, K.~Zhou, H.~Hao, Y.~Zhao, T.~Zhang, S.~Gao, and J.~Liu,
  ``Ce-net: Context encoder network for 2d medical image segmentation,''
  \emph{{IEEE} Trans. Med. Imag.}, vol.~38, no.~10, pp. 2281--2292, 2019.

\bibitem{b48}
S.~Lal, D.~Das, K.~Alabhya, A.~Kanfade, A.~Kumar, and J.~Kini, ``Nucleisegnet:
  robust deep learning architecture for the nuclei segmentation of liver cancer
  histopathology images,'' \emph{Comput. Biol. Med.}, vol. 128, p. 104075,
  2021.

\bibitem{b49}
S.~E.~A. Raza, L.~Cheung, M.~Shaban, S.~Graham, D.~Epstein, S.~Pelengaris,
  M.~Khan, and N.~M. Rajpoot, ``Micro-net: A unified model for segmentation of
  various objects in microscopy images,'' \emph{Med. Image Anal.}, vol.~52, pp.
  160--173, 2019.

\bibitem{b50}
Z.~Zhou, M.~M.~R. Siddiquee, N.~Tajbakhsh, and J.~Liang, ``Unet++: Redesigning
  skip connections to exploit multiscale features in image segmentation,''
  \emph{{IEEE} Trans. Med. Imag.}, vol.~39, no.~6, pp. 1856--1867, 2019.

\bibitem{b51}
P.~Quelhas, M.~Marcuzzo, A.~M. Mendon{\c{c}}a, and A.~Campilho, ``Cell nuclei
  and cytoplasm joint segmentation using the sliding band filter,''
  \emph{{IEEE} Trans. Med. Imag.}, vol.~29, no.~8, pp. 1463--1473, 2010.


\end{thebibliography}
\end{document}